\documentclass[a4paper,fleqn,usenatbib,useAMS]{mnras}
\usepackage{graphicx}
\usepackage{ltxtable}

\def\e{{\rm e}}

\def\lesssim{\mathrel{\hbox{\rlap{\hbox{\lower4pt\hbox{$\sim$}}}\hbox{$<$}}}}
\def\gtrsim{\mathrel{\hbox{\rlap{\hbox{\lower4pt\hbox{$\sim$}}}\hbox{$>$}}}}

\title{Hyperfine transitions of $^{13}$CN from pre-protostellar sources }
\author[D. R. Flower et al.]{D. R. Flower$^{1}$\thanks{E-mail:
david.flower@durham.ac.uk}, P. Hily-Blant$^{2}$ \\
$^1$Physics Department, The University, Durham DH1 3LE, UK \\
$^2$LAOG (UMR 5571), Universit\'{e} de Grenoble, BP 53, F-38041 Grenoble Cedex 09, France \\}

\begin{document}

\date{Accepted 2008 December 15. Received 2008 December 14; in original form 2008 October 11}

\pagerange{\pageref{firstpage}--\pageref{lastpage}} \pubyear{2008}

\maketitle

\label{firstpage}

\begin{abstract}

Recent quantum mechanical calculations of rate coefficients for collisional transfer of population between the hyperfine states of $^{13}$CN enable their population densities to be determined. We have computed the relative populations of the hyperfine states of the $N = 0, 1, 2$ rotational states for kinetic temperatures $5 \le T \le 20$~K and molecular hydrogen densities  $1 \le n$(H$_2$) $\le 10^{10}$~cm$^{-3}$.  Spontaneous and induced radiative transitions were taken into account. Our calculations show that, if the lines are optically thin, the populations of the hyperfine states, $F$, within a given rotational manifold are proportional to their statistical weights, $(2F + 1)$ -- i.e. in {\sc LTE} -- over the entire range of densities. We have re-analyzed {\sc IRAM} 30~m telescope observations of $^{13}$CN hyperfine transitions ($N = 1 \rightarrow 0$) in four starless cores. A comparison of these observations with our calculations confirms that the hyperfine states are statistically populated in these sources.

\end{abstract}


%

\section{Introduction}
\label{sec:intro}

The interpretation of the emission lines of molecules in the interstellar medium ({\sc ISM}) is often complicated by the effects of re-absorption and scattering, owing to significant optical depths in the lines. Partly for this reason, observations of less abundant isotopologues are analyzed, in addition or in preference to those of the principal species; this is the case of CO, for example, where $^{13}$CO and C$^{18}$O lines are used, and also of CN, where $^{13}$CN and C$^{15}$N serve a similar purpose. In the present paper, we consider the emission lines of $^{13}$CN, observed at millimetre wavelengths in pre-protostellar sources.

$^{13}$CN has a rich spectrum at mm-wavelengths, where it displays the effects of the fine structure interaction, between the electron spin and the nuclear rotation, and of the hyperfine interaction with the spins of the $^{13}$C ($I_1 = \frac {1}{2}$) and N ($I_2 = 1$) nuclei. Thus, Bogey et al. (1984) listed 16 `allowed' transitions, in the vicinity of 100~GHz, between the rotational states $N = 0$ and $N = 1$, and a further 25 transitions at 200~GHz between the $N = 1$ and $N = 2$ states. It is this cornucopia of optically thin transitions that one wishes to exploit, in order to obtain a better understanding of the conditions in pre-protostellar objects.

It is generally assumed that the hyperfine levels, $F$, of a given $N$ are populated in proportion to their statistical weights, $(2F + 1)$, i.e. that they are in local thermodynamic equilibrium {\sc LTE}. However, {\sc LTE} is the exception, rather than the rule, in the {\sc ISM}, because very low densities prevail. The assumption of {\sc LTE} is usually dictated by a lack of rate coefficients for collisional population transfer between the hyperfine levels of the molecule. In the case of $^{13}$CN (and of C$^{15}$N), this situation has been rectified recently by the calculations of Flower \& Lique (2015), which provide rate coefficients for collisions with para-H$_2$ in its rotational ground state -- the dominant perturber at low kinetic temperatures, $T$. Thus, the opportunity arises to calculate the hyperfine level populations explicitly, allowing for collisional and radiative transfer, where the latter is induced by the cosmic microwave background or by the emission of dust present in the medium.

We present the details of the calculations in Section~\ref{calc} and the numerical results and their implications in the following Section~\ref{results}. Our concluding remarks are in Section~\ref{conclusions}.

\section{Calculations}
\label{calc}

Bogey et al. (1984) note that the rotational ($N$) and the hyperfine ($F$) are the `good' quantum numbers of $^{13}$CN. Accordingly, we evaluate the relative populations, in equilibrium, of the levels ($N, F$), for a wide range of values of the density of molecular hydrogen (specifically, para-H$_2$), $1 \le n$(H$_2$) $\le 10^{10}$~cm$^{-3}$, and a range of kinetic temperatures, $5 \le T \le 20$~K. Collisional and radiative population transfer are taken into account, as described below.

\subsection{Collisional population transfer}
\label{coll}

We use the rate coefficients for collisions between $^{13}$CN and para-H$_2$ that were calculated by Flower \& Lique (2015), who used the angular momentum coupling scheme $\bf {j } = \bf{N} + \bf{S}$, $\bf {I} = \bf{I}_1 + \bf{I}_2$, $\bf {F} = \bf{j} + \bf{I}$, where $N$ is the rotational quantum number, $S = \frac{1}{2}$ is the net electron spin, $I_1 = \frac {1}{2}$ and $I_2 = 1$ are the net spin quantum numbers of the $^{13}$C and N nuclei, respectively, and $F$ is the hyperfine quantum number. Their calculations were performed separately for the cases of $I = \frac {1}{2}$ and $I = \frac {3}{2}$, as the value of the total nuclear spin, $I$, is conserved during the collision. For a given value of $I$, the rate (s$^{-1}$) of collisional transfer of population from the state ($N, j, F$) to the state ($N^{\prime}, j^{\prime}, F^{\prime}$) is $n$(H$_2$)$q(N, j, F \rightarrow N^{\prime}, j^{\prime}, F^{\prime})$, and hence the total rate of transfer from level ($N, F$) to level ($N^{\prime}, F^{\prime}$) is $n$(H$_2$)$\sum _{j, j^{\prime}} q(N, j, F \rightarrow N^{\prime}, j^{\prime}, F^{\prime})$, where $q$ denotes a rate coefficient (cm$^3$~s$^{-1}$) and $n$(H$_2$) is the density (cm$^{-3}$) of ground state para-H$_2$. We assume that the nuclear spin states $I = \frac {1}{2}$ and $I = \frac {3}{2}$ are statistically populated and evaluate the rate coefficients as

\begin{displaymath}
q(N, j, F \rightarrow N^{\prime}, j^{\prime}, F^{\prime}) = \frac {1}{3} q(I = \frac {1}{2}|N, j, F \rightarrow N^{\prime}, j^{\prime}, F^{\prime}) +
\end{displaymath}
\begin{equation}
 \frac {2}{3} q(I = \frac {3}{2}|N, j, F \rightarrow N^{\prime}, j^{\prime}, F^{\prime}).
\label{equ1}
\end{equation}
Varying the relative populations of the nuclear spin states had no significant effect on the results of the calculations reported in Section~\ref{results} below.

\subsection{Radiative population transfer}
\label{rad}

Following the analysis of optically allowed transitions by Nicholls \& Stewart (1962), the rate (s$^{-1}$) of spontaneous radiative transitions between states $i$ and $j$, belonging to levels $k$ and $l$, respectively, with energies $E_k < E_l$, is given by

\begin{equation}
A_{j i} = \frac {4 (E_l - E_k)^3}{3 \hbar ^4 c^3} S_{j i},
\label{equ2}
\end{equation}
where the line strength

\begin{equation}
S_{j i} = s_{j i} \mu ^2,
\label{equ3}
\end{equation}
and where $\mu = 1.45$~debye (cf. Hily-Blant et al. 2013) is the electric dipole moment of the molecule in its vibrational ground state. The algebraic quantities $s_{j i}$ were computed by Bogey et al. (1984, table~3) for dipole transitions within the $N = 0 \leftarrow 1$ and $N = 1 \leftarrow 2$ rotational manifolds. We note that

\begin{equation}
S_{l k} = \sum _{i, j} S_{j i}.
\label{equ4}
\end{equation}
In the present context, the levels $k, l$ are identified by the quantum numbers ($N, F$), and the states $i, j$ by ($N, F_1, F_2, F$), where $\bf {F_1 } = \bf{S} + \bf{I_1}$, $\bf {F}_2 = \bf{N} + \bf{F}_1$, $\bf {F} = \bf{F}_2 + \bf{I}_2$ is the angular momentum coupling scheme adopted by Bogey et al. (1984). The $A$-values for transitions between pairs of these states are given in Table~\ref{tab1}. The accuracy of these results is determined principally by the accuracy of the dipole moment, which is difficult to assess from the available data (http://cccbdb.nist.gov/dipole2.asp). The rates (s$^{-1}$) of spontaneous transitions between levels ($N, F$) were obtained by implementing the summation relation~(\ref{equ4}) above.

According to the Einstein relations, the rates of radiative excitation and de-excitation by a radiation field of energy density $\rho _{\nu } {\rm d}\nu $ (erg~cm$^{-3}$) at frequency $\nu $ are $B_{kl} \rho _{\nu }$ and $B_{lk} \rho _{\nu }$, respectively, where 

\begin{equation}
\omega _k B_{kl} = \omega _l B_{l k} = \frac {c^3}{8\pi h\nu ^3} \omega _l A_{l k}
\label{equ5}
\end{equation}
and $\omega = (2F + 1)$ is the statistical weight. We adopt a black--body radiation field,

\begin{equation}
\rho _{\nu } (T_{\rm b}) = \frac {8\pi h\nu ^3}{c^3} \frac {1}{\exp {\left (\frac {h\nu }{kT_{\rm b}}\right )} - 1}
\label{equ6}
\end{equation}
where $T_{\rm b} = 2.73$~K is the temperature of the cosmic background radiation ({\sc CBR}) or of a black--body radiation field produced by local dust, mixed with the gas.

\subsection{Level populations and line intensities}
\label{pops}

Knowing the total rates, 

\begin{equation}
p_{kl} = n({\rm H}_2)q_{kl}(T) + A_{kl} + B_{kl},
\label{equ6.1}
\end{equation}
of collisional and radiative population transfer between levels ($N, F$), the populations, $n_k$, of the 12 levels with rotational quantum numbers $N = 0, 1, 2$ may be obtained by solving the rate equations

\begin{equation}
\frac {{\rm d}n_k}{{\rm d}t} = \sum _{l \ne k} (n_l p_{lk} - n_k p_{kl}) 
\label{equ7}
\end{equation}
together with the closure relation

\begin{equation}
\sum _{k} n_k = n(^{13}{\rm CN}),
\label{equ8}
\end{equation}
where $n(^{13}{\rm CN})$ is the number density of $^{13}$CN molecules. In steady--state, $\frac {{\rm d}n_k}{{\rm d}t} \equiv 0$, and the relative level populations are readily obtained by matrix inversion, setting $n(^{13}{\rm CN}) = 1$. Then, the relative line emission (erg~s$^{-1}$) is given by

\begin{equation}
I_{ji} = n_l A_{ji} h\nu _{ji},
\label{equ9}
\end{equation}
where, we recall, $j$ and $i$ are states ($N, F_1, F_2, F$) belonging to the levels $l$ and $k$, identified by the quantum numbers ($N, F$); $\nu _{ji}$ is the line frequency.

\begin{table*}
\caption[]{Spontaneous radiative transitions probabilities, $A$, and relative line emission, $I$ [equ.~(\ref {equ9})], for transitions $N = 0 \leftarrow N^{\prime } = 1$ and $N = 1 \leftarrow N^{\prime } = 2$; $T = 10$~K and $n$(H$_2$) = $10^5$~cm$^{-3}$. The angular momentum coupling scheme of Bogey et al. (1984) is adopted. The quantity $s$ is the algebraic factor in the expression for the line strength, equ.~(\ref {equ3}), and $\nu $ is the line frequency. Note that the degeneracy of the upper (emitting) level is $(2F^{\prime} + 1)$. Numbers in parentheses are powers of 10.}

\begin{center}
\begin{tabular}{rrrrrrrr}
\hline
           \noalign{\smallskip}
      $N$ & $F_1$ & $F_2 \leftarrow F_2^{\prime }$ & $F \leftarrow F^{\prime }$ &  $s$  & $\nu $ MHz & $A$ s$^{-1}$ & $I$ erg~s$^{-1}$ \\
           \noalign{\smallskip}
\hline
           \noalign{\smallskip}
      0  &   0 &   0 $\leftarrow $ 1  &   1 $\leftarrow $ 2  &   1.66 &   108651.297 & 1.04(-5) & 1.39(-1) \\
      0  &   0 &   0 $\leftarrow $ 1  &   1 $\leftarrow $ 1  &   1.00 &   108636.923 & 1.05(-5) & 8.57(-2) \\
      0  &   0 &   0 $\leftarrow $ 1  &   1 $\leftarrow $ 0  &   0.33 &   108631.121 & 1.04(-5) & 2.81(-2) \\
      0  &   1 &   1 $\leftarrow $ 2  &   0 $\leftarrow $ 1  &   0.56 &   108786.982 & 5.88(-6) & 4.80(-2) \\
      0  &   1 &   1 $\leftarrow $ 2  &   1 $\leftarrow $ 2  &   1.25 &   108782.374 & 7.87(-6) & 1.05(-1) \\
      0  &   1 &   1 $\leftarrow $ 2  &   2 $\leftarrow $ 3  &   2.33 &   108780.201 & 1.05(-5) & 1.94(-1) \\
      0  &   1 &   1 $\leftarrow $ 2  &   1 $\leftarrow $ 1  &   0.42 &   108793.753 & 4.41(-6) & 3.60(-2) \\
      0  &   1 &   1 $\leftarrow $ 2  &   2 $\leftarrow $ 2  &   0.42 &   108796.400 & 2.65(-6) & 3.54(-2) \\
      0  &   1 &   1 $\leftarrow $ 1  &   0 $\leftarrow $ 1  &   0.33 &   108638.212 & 3.45(-6) & 2.81(-2) \\
      0  &   1 &   1 $\leftarrow $ 1  &   1 $\leftarrow $ 2  &   0.42 &   108643.590 & 2.64(-6) & 3.53(-2) \\
      0  &   1 &   1 $\leftarrow $ 1  &   1 $\leftarrow $ 1  &   0.25 &   108645.064 & 2.62(-6) & 2.14(-2) \\
      0  &   1 &   1 $\leftarrow $ 1  &   2 $\leftarrow $ 2  &   1.25 &   108657.646 & 7.85(-6) & 1.05(-1) \\
      0  &   1 &   1 $\leftarrow $ 1  &   1 $\leftarrow $ 0  &   0.33 &   108644.346 & 1.04(-5) & 2.81(-2) \\
      0  &   1 &   1 $\leftarrow $ 1  &   2 $\leftarrow $ 1  &   0.42 &   108658.948 & 4.39(-6) & 3.58(-2) \\
      0  &   1 &   1 $\leftarrow $ 0  &   1 $\leftarrow $ 1  &   0.33 &   108412.862 & 3.43(-6) & 2.79(-2) \\
      0  &   1 &   1 $\leftarrow $ 0  &   2 $\leftarrow $ 1  &   0.56 &   108426.889 & 5.82(-6) & 4.74(-2) \\
           \noalign{\smallskip}
      1  &   0 &   1 $\leftarrow $ 2  &   0 $\leftarrow $ 1  &   0.33 &   217296.605 & 2.76(-5) & 3.00(-2) \\
      1  &   0 &   1 $\leftarrow $ 2  &   1 $\leftarrow $ 2  &   0.75 &   217301.175 & 3.77(-5) & 6.25(-2) \\
      1  &   0 &   1 $\leftarrow $ 2  &   2 $\leftarrow $ 3  &   1.40 &   217303.191 & 5.02(-5) & 1.19(-1) \\
      1  &   0 &   1 $\leftarrow $ 2  &   1 $\leftarrow $ 1  &   0.25 &   217290.823 & 2.09(-5) & 2.27(-2) \\
      1  &   0 &   1 $\leftarrow $ 2  &   2 $\leftarrow $ 2  &   0.25 &   217286.804 & 1.26(-5) & 2.09(-2) \\
      1  &   1 &   2 $\leftarrow $ 3  &   1 $\leftarrow $ 2  &   0.84 &   217469.151 & 4.23(-5) & 7.02(-2) \\
      1  &   1 &   2 $\leftarrow $ 3  &   2 $\leftarrow $ 3  &   1.24 &   217467.150 & 4.46(-5) & 1.06(-1) \\
      1  &   1 &   2 $\leftarrow $ 3  &   3 $\leftarrow $ 4  &   1.80 &   217467.150 & 5.03(-5) & 1.61(-1) \\
      1  &   1 &   2 $\leftarrow $ 3  &   2 $\leftarrow $ 2  &   0.16 &   217480.559 & 8.05(-6) & 1.34(-2) \\
      1  &   1 &   2 $\leftarrow $ 3  &   3 $\leftarrow $ 3  &   0.16 &   217483.606 & 5.75(-6) & 1.37(-2) \\
      1  &   1 &   1 $\leftarrow $ 2  &   0 $\leftarrow $ 1  &   0.25 &   217443.722 & 2.10(-5) & 2.29(-2) \\
      1  &   1 &   1 $\leftarrow $ 2  &   1 $\leftarrow $ 2  &   0.57 &   217436.350 & 2.87(-5) & 4.76(-2) \\
      1  &   1 &   1 $\leftarrow $ 2  &   2 $\leftarrow $ 3  &   1.05 &   217428.563 & 3.77(-5) & 8.96(-2) \\
      1  &   1 &   1 $\leftarrow $ 2  &   1 $\leftarrow $ 1  &   0.19 &   217443.722 & 1.59(-5) & 1.73(-2) \\
      1  &   1 &   1 $\leftarrow $ 2  &   2 $\leftarrow $ 2  &   0.19 &   217437.702 & 9.56(-6) & 1.59(-2) \\
      1  &   1 &   2 $\leftarrow $ 2  &   1 $\leftarrow $ 1  &   0.11 &   217294.470 & 9.21(-6) & 1.00(-2) \\
      1  &   1 &   2 $\leftarrow $ 2  &   2 $\leftarrow $ 2  &   0.17 &   217298.937 & 8.54(-6) & 1.42(-2) \\
      1  &   1 &   2 $\leftarrow $ 2  &   3 $\leftarrow $ 3  &   0.31 &   217306.117 & 1.11(-5) & 2.64(-2) \\
      1  &   1 &   0 $\leftarrow $ 1  &   1 $\leftarrow $ 2  &   0.56 &   217304.927 & 2.81(-5) & 4.66(-2) \\
      1  &   1 &   0 $\leftarrow $ 1  &   1 $\leftarrow $ 1  &   0.33 &   217277.680 & 2.76(-5) & 3.00(-2) \\
      1  &   1 &   0 $\leftarrow $ 1  &   1 $\leftarrow $ 0  &   0.11 &   217264.639 & 2.76(-5) & 1.00(-2) \\
      1  &   1 &   1 $\leftarrow $ 1  &   1 $\leftarrow $ 2  &   0.10 &   217072.801 & 5.01(-6) & 8.29(-3) \\
      1  &   1 &   1 $\leftarrow $ 1  &   2 $\leftarrow $ 2  &   0.31 &   217074.239 & 1.55(-5) & 2.57(-2) \\
      1  &   1 &   1 $\leftarrow $ 1  &   1 $\leftarrow $ 0  &   0.08 &   217032.603 & 2.00(-5) & 7.27(-3) \\
      1  &   1 &   1 $\leftarrow $ 1  &   2 $\leftarrow $ 1  &   0.10 &   217046.988 & 8.34(-6) & 9.06(-3) \\
\hline
\end{tabular}

\end{center}
\label{tab1}
\end{table*}

\begin{table*}
\caption[]{Integrated intensities ($T{\rm d}v$ mK~km~s$^{-1}$; main--beam temperature scale) from Gaussian fits to hfs lines $N = 0 \leftarrow N^{\prime } = 1$, $F_1 = 1$, observed in the four outflow sources listed, at the offsets (d$x\arcsec, $ d$y\arcsec $) indicated. The error bars are $1\sigma$ .}

\begin{center}
\begin{tabular}{rrrrrrrr}
\hline
           \noalign{\smallskip}
      $F_2 \leftarrow F_2^{\prime }$ & $F \leftarrow F^{\prime }$ & $\nu $ MHz & L1544(0, 0) &  L183(0, 0) & L1517B(-10, -20) & Oph~D(0, 0) \\
           \noalign{\smallskip}
\hline
           \noalign{\smallskip}
      1 $\leftarrow $ 2  &   2 $\leftarrow $ 3  &   108780.201 & $73.5\pm 3.9$ & $32.0\pm 3.0$ & $31.3\pm 2.9$ & $25.7\pm 6.8$ \\
      1 $\leftarrow $ 2  &   1 $\leftarrow $ 2  &   108782.374 & $43.2\pm 3.8$ & $12.0\pm 3.0$ & $13.7\pm 2.0$ &  $11.9\pm 2.5$ \\
      1 $\leftarrow $ 2  &   0 $\leftarrow $ 1  &   108786.982 & $19.3\pm 4.0$ & $10.0\pm 2.0$ & $9.3\pm 3.5$ &  $6.5\pm 2.1$ \\
      1 $\leftarrow $ 2  &   1 $\leftarrow $ 1  &   108793.753 & $19.3\pm 4.5$ & $0.0\pm 0.0$ &  $7.6\pm 2.4$ & $5.1\pm 3.6$ \\
      1 $\leftarrow $ 2  &   2 $\leftarrow $ 2  &   108796.400 & $18.9\pm 3.3$ & $0.8\pm 0.9$ &  $5.2\pm 2.0$ & $6.0\pm 3.1$ \\
           \noalign{\smallskip}
\hline
\end{tabular}

\end{center}
\label{tab2}
\end{table*}

\begin{table*}
\caption[]{Relative intensities of hfs lines $N = 0 \leftarrow N^{\prime } = 1$, $F_1 = 1$, observed in the four starless cores listed, at the offsets (d$x\arcsec, $ d$y\arcsec $) indicated. The values calculated from the results in Table~\ref{tab1} are also given.}

\begin{center}
\begin{tabular}{rrrrrrrr}
\hline
           \noalign{\smallskip}
      $F_2 \leftarrow F_2^{\prime }$ & $F \leftarrow F^{\prime }$ & $\nu $ MHz & L1544(0, 0) &  L183(0, 0) & L1517B(-10, -20) & Oph~D(0, 0) & Calc. \\
           \noalign{\smallskip}
\hline
           \noalign{\smallskip}
      1 $\leftarrow $ 2  &   2 $\leftarrow $ 3  &   108780.201 & $1.00$ & $1.00$ & $1.00$ & $1.00$ & $1.00$ \\
      1 $\leftarrow $ 2  &   1 $\leftarrow $ 2  &   108782.374 & $0.59\pm 0.06$ & $0.38\pm 0.10$ & $0.44\pm 0.08$ &  $0.46\pm 0.16$ & 0.54 \\
      1 $\leftarrow $ 2  &   0 $\leftarrow $ 1  &   108786.982 & $0.26\pm 0.06$ & $0.31\pm 0.07$ & $0.30\pm 0.11$ &  $0.25\pm 0.10$ & 0.25 \\
      1 $\leftarrow $ 2  &   1 $\leftarrow $ 1  &   108793.753 & $0.26\pm 0.06$ & $0.00\pm 0.00$ &  $0.24\pm 0.08$ & $0.20\pm 0.15$ & 0.19 \\
      1 $\leftarrow $ 2  &   2 $\leftarrow $ 2  &   108796.400 & $0.26\pm 0.05$ & $0.03\pm 0.03$ &  $0.16\pm 0.07$ & $0.23\pm 0.14$ & 0.18 \\
           \noalign{\smallskip}
\hline
\end{tabular}

\end{center}
\label{tab3}
\end{table*}

\subsection{Line opacities}
\label{opac}

The opacity at frequency $\nu $ in a transition of oscillator strength $f_{i j}$ is given by

\begin{equation}
\kappa_{i j}(\nu ) = n(^{13}{\rm CN}) \frac {n_i}{n(^{13}{\rm CN})} \frac {\pi e^2}{m c}f_{i j}\phi (\nu ),
\label{equ10}
\end{equation}
where $m$ is the mass of the electron and $\phi (\nu )$ is the line profile function; the spontaneous radiative transition probability and the absorption oscillator strength are related by

\begin{equation}
\omega_{j} A_{j i} = \frac {8\pi ^2 e^2 \nu ^2}{m c^3} \omega_{i} f_{i j}
\label{equ11}
\end{equation}
(Nicholls \& Stewart 1962). If the line profile is taken to be Gaussian, then

\begin{equation}
\phi (\nu ) = \frac {1}{\nu _0} \left (\frac {\beta } {\pi } \right )^{0.5} \exp {- \frac {\beta (\nu - \nu _0)^2}{\nu _0^2}},
\label{equ12}
\end{equation}
where $\beta = M c^2 / (2 k T_{\rm D})$, $M$ is the mass of the $^{13}$CN molecule and $T_{\rm D}$ is the corresponding Doppler temperature; $\nu _0$ is the frequency at the centre of the line. If $\Delta \nu $ is defined as the $(1/\e )$ half--width of the line, then $\phi (\nu _0) = 1/(\sqrt {\pi } \Delta \nu )$.

Let us consider the strongest of the $^{13}$CN hyperfine transitions, at a frequency of 108780.201~MHz. From (\ref{equ11}) and Table~\ref{tab1}, the absorption oscillator strength of this transition is $f_{i j} = 1.68 \times 10^{-6}$. Integrating equ.~(\ref{equ10}) over the line of sight, and assuming that the fractional level population, $\frac {n_i}{n(^{13}{\rm CN})}$, is constant, the optical depth at the line centre is

\begin{displaymath}
\tau (\nu _0) = N(^{13}{\rm CN}) \frac {n_i}{n(^{13}{\rm CN})} \frac {\pi e^2}{m c}f_{i j} \frac {1}{\sqrt {\pi } \Delta \nu },
\end{displaymath}
where $N(^{13}{\rm CN})$ is the column density of the molecule.

In L1544 and L183, Hily-Blant et al. (2008, tables 1 and 2) list column densities $N(^{13}{\rm CN}) \approx 10^{12}$~cm$^{-2}$ at offset (0, 0); from their figs.~C.1 and C.2, $\Delta \nu \approx 0.2$~km~s$^{-1}$, equivalent to $\Delta \nu \approx 0.07$~MHz at the line frequency. Our calculations yield $n_i / n(^{13}{\rm CN}) \approx 10^{-1}$. It follows that $\tau (\nu _0) \approx 10^{-2}$. For column densities $N(^{13}{\rm CN}) \gtrsim 10^{14}$~cm$^{-2}$, $\tau (\nu _0) \gtrsim 1$, and deviations from the LTE line ratios are to be expected. Departures of hyperfine level populations from {\sc LTE} have been attributed to line--optical--depth effects in the cases of N$_2$H$^+$ (Daniel et al. 2006) and, possibly, NH$_2$D (Daniel et al. 2013).

In the following Section~\ref{results}, we assume the $^{13}$CN hyperfine transitions to be optically thin and calculate their intensities following Section~\ref{pops}.

\section{Results and discussion}
\label{results}

\begin{figure}
  \centering
  \includegraphics[width=\hsize]{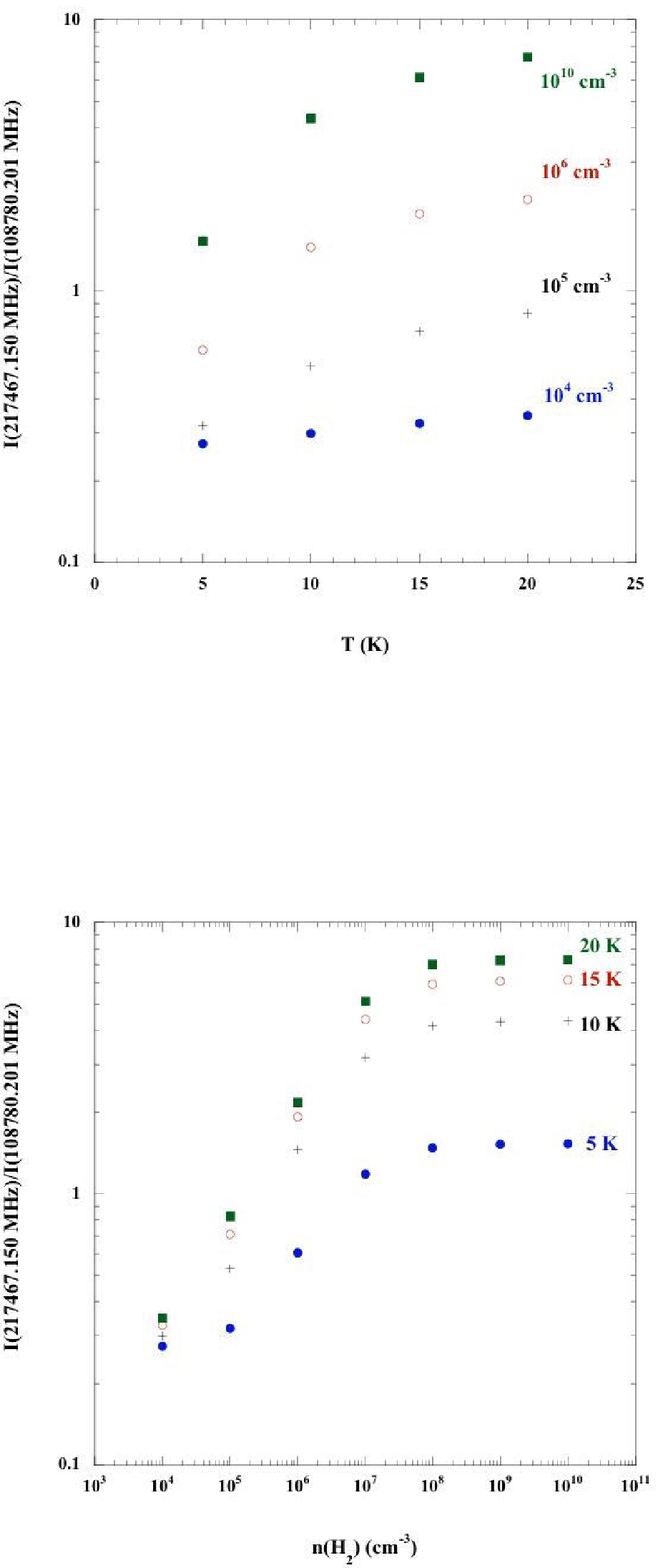}
  \caption{The relative intensity, I(217467.150 MHz)/I(108780.201
    MHz), of the strongest of the transitions between $N= 2$ and $N =
    1$ (217467.150 MHz) and between $N= 1$ and $N = 0$ (108780.201
    MHz); the relative intensity is plotted as a function of both
    temperature, $T$ (upper panel), and para-H$_2$ density, $n$(H$_2$)
    (lower panel).}
  \label{fig1}
\end{figure}

The relative level populations and line intensities have been computed for a range of values of the kinetic temperature, $T$, and the para-H$_2$ density, $n$(H$_2$). In Table \ref{tab1} are listed the values of the relative line emission, $I$ [equ.~(\ref {equ9})], determined separately for transitions within the $N = 0 \leftarrow N^{\prime } = 1$ and $N = 1 \leftarrow N^{\prime } = 2$ rotational manifolds and for $T = 10$~K and $n$(H$_2$) = $10^5$~cm$^{-3}$, values typical of pre-protostellar cores; the {\sc CBR} was included in these calculations.

The values of the relative line emission in Table~\ref{tab1} for $N = 0 \leftarrow N^{\prime } = 1$ differ insignificantly from the corresponding results in table~A.1 of Hily-Blant et al. (2008), calculated in {\sc LTE} -- i.e. assuming that the populations of the hyperfine states of given $N$ are in proportion to their statistical weights, $(2F + 1)$. In fact, we found negligible variations of the relative line emission within a rotational manifold as $n$(H$_2$) was varied from the low to the high density limit, including either the {\sc CBR} or a black--body radiation field that is assumed to be emitted by dust grains in and at the same temperature as the gas. This result reflects the fact that neither collisional nor radiative transitions discriminate between the hyperfine states, $F$, with a given value of $N$, and that each magnetic sub-state, $M_F$, is equally populated -- a rare example of {\sc LTE} conditions prevailing in the {\sc ISM}. Whilst it has long been assumed that {\sc LTE} applies to the relative populations of molecular hyperfine levels (cf. Walmsley et al. 1982),  the present study is the first to demonstrate the validity of this assumption, in the case of $^{13}$CN. An analogous conclusion was reached by Daniel et al. (2006) regarding N$_2$H$^+$, in the optically thin regime.

Unlike the hyperfine transitions of CN and HCN, which have been considered by Faure \& Lique (2012), the lines of $^{13}$CN are likely to be optically thin (cf. Section~\ref{opac}), given that the isotopic abundance ratio $^{13}$C:$^{12}$C $\approx $ 1:70 in the {\sc ISM} (Milam et al. 2005). It is surprising, therefore, that Hily-Blant et al. (2010) concluded that the relative intensities of some of the $N = 1 \rightarrow 0$ hyperfine transitions of $^{13}$CN, observed in four starless cores, depart significantly from {\sc LTE}. However, a close scrutiny of their fig.~3 reveals that the relative ``{\sc LTE}'' values in this figure are inconsistent with the correct values, given by Hily-Blant et al. (2008, table~B.2). 

We have re-analyzed our {\sc IRAM} 30~m telescope observations of the four starless cores L1544, L183, L1517B, and Oph~D, and the integrated line intensities -- at the offset for which the lines are strongest in each source -- are reported in Table~\ref{tab2}; the intensities relative to the strongest transition, at 108780.201~MHz, are given in Table~\ref{tab3}, together with the corresponding theoretical results from Table~\ref{tab1}. Our conclusion from this comparison is that the relative populations of the hyperfine levels of $^{13}$CN within a given rotational state, $N$, do not depart significantly from {\sc LTE} in these sources.

Although {\sc LTE} is found to apply to the relative populations of hyperfine levels of a given rotational state, $N$, the same is not true of the relative populations of levels belonging to different values of $N$. This situation is reminiscent of that which obtains in molecules such as H$_2$, where the relative populations of the rotational levels within a given vibrational manifold can attain {\sc LTE} at densities which are much lower than are required to thermalize the relative populations of different vibrational states. In Fig.~\ref{fig1} is plotted the relative intensity of the strongest of the transitions ($F_1 = 1$; $F_2 \leftarrow F_2^{\prime }$, $F \leftarrow F^{\prime }$) between $N= 2$ and $N = 1$ ($2 \leftarrow  3$, $3 \leftarrow  4$) and between $N= 1$ and $N = 0$ ($1 \leftarrow  2$, $2 \leftarrow  3$). Results are given as functions of both temperature, $T$, and para-H$_2$ density, $n$(H$_2$); {\sc LTE} is approached at densities $n$(H$_2$)$ \gtrsim 10^{7}$~cm$^{-3}$, and $n$(H$_2$)$ = 10^{10}$~cm$^{-3}$ is effectively the high density limit, at which a Boltzmann distribution of population obtains. As may be seen from Fig.~\ref{fig1}, the relative line intensity is much more sensitive to $T$ than to $n$(H$_2$), for the relevant ranges of these parameters, which opens the possibility of using this line ratio as a thermometer in starless cores.

\section{Concluding remarks}
\label{conclusions}

We have computed the relative population densities of the hyperfine states that comprise the $N = 0, 1, 2$ rotational manifolds of $^{13}$CN, in the range of kinetic temperature $5 \le T \le 20$~K and of molecular hydrogen density  $1 \le n$(H$_2$) $\le 10^{10}$~cm$^{-3}$. The rate coefficients for collisional transfer between the hyperfine states derive from the recent quantum mechanical calculations of Flower \& Lique (2015), which relate specifically to $^{13}$CN--para-H$_2$. Spontaneous radiative transitions, and those induced by the {\sc CBR}, were taken into account. We find that the hyperfine states, $F$, with a given value of the rotational quantum number, $N$, are populated in proportion to their statistical weights, $(2F + 1)$, as expected in {\sc LTE}, even at densities that are much lower than required to thermalize the populations of levels with differing values of $N$. A re-analysis of observations of hyperfine transitions $N = 1 \rightarrow 0$ of $^{13}$CN, observed in a number of pre-protostellar sources, confirms that the hyperfine states are, indeed, statistically populated.

\section*{Acknowledgments}

DRF acknowledges support from STFC (ST/L00075X/1), including provision of local computing resources.

\bibliographystyle{mnras}

\begin{thebibliography}{}
\makeatletter
\relax
\def\mn@urlcharsother{\let\do\@makeother \do\$\do\&\do\#\do\^\do\_\do\%\do\~}
\def\mn@doi{\begingroup\mn@urlcharsother \@ifnextchar [ {\mn@doi@}
  {\mn@doi@[]}}
\def\mn@doi@[#1]#2{\def\@tempa{#1}\ifx\@tempa\@empty \href
  {http://dx.doi.org/#2} {doi:#2}\else \href {http://dx.doi.org/#2} {#1}\fi
  \endgroup}
\def\mn@eprint#1#2{\mn@eprint@#1:#2::\@nil}
\def\mn@eprint@arXiv#1{\href {http://arxiv.org/abs/#1} {{\tt arXiv:#1}}}
\def\mn@eprint@dblp#1{\href {http://dblp.uni-trier.de/rec/bibtex/#1.xml}
  {dblp:#1}}
\def\mn@eprint@#1:#2:#3:#4\@nil{\def\@tempa {#1}\def\@tempb {#2}\def\@tempc
  {#3}\ifx \@tempc \@empty \let \@tempc \@tempb \let \@tempb \@tempa \fi \ifx
  \@tempb \@empty \def\@tempb {arXiv}\fi \@ifundefined
  {mn@eprint@\@tempb}{\@tempb:\@tempc}{\expandafter \expandafter \csname
  mn@eprint@\@tempb\endcsname \expandafter{\@tempc}}}

\makeatother
\end{thebibliography}


\begin{thebibliography}{plainnat}
\bibitem[\protect\citeauthoryear{Author}{2013}]{}Bogey, M., Demuynck, C., Destombes, J. L. 1984, Can. J. Phys., 62, 1248
\bibitem[\protect\citeauthoryear{Author}{2013}]{}Daniel, F., Cernicharo, J., Dubernet, M.-L.  2006, ApJ, 648, 461
\bibitem[\protect\citeauthoryear{Author}{2013}]{}Daniel, F., G\'{e}rin, M., Roueff, E., Cernicharo, J., Marcelino, N., Lique, F., Lis, D. C., Teyssier, D., Biver, N., Bockel\'{e}e-Morvan, D. 2013, A\&A, 560, A3
\bibitem[\protect\citeauthoryear{Author}{2013}]{}Faure, A., Lique, F. 2012, MNRAS, 425, 740
\bibitem[\protect\citeauthoryear{Author}{2013}]{}Flower, D. R., Lique, F. 2015, MNRAS, 446, 1750  
\bibitem[\protect\citeauthoryear{Author}{2013}]{}Hily-Blant, P., Walmsley, C. M., Pineau des For\^ets, G., Flower, D. R. 2008, A\&A, 480, L5
\bibitem[\protect\citeauthoryear{Author}{2013}]{}Hily-Blant, P., Walmsley, C. M., Pineau des For\^ets, G., Flower, D. R. 2010, A\&A, 513, A41
\bibitem[\protect\citeauthoryear{Author}{2013}]{}Hily-Blant, P., Pineau des For\^ets, G., Faure, A., Le Gal, R., Padovani, M. 2013, A\&A, 557, A65
\bibitem[\protect\citeauthoryear{Author}{2013}]{}Milam, S. N., Savage, C., Brewster, M. A., Ziurys, L. M., Wyckoff, S. 2005, ApJ, 634, 1126
\bibitem[\protect\citeauthoryear{Author}{2013}]{}Nicholls, R. W., Stewart, A. L. 1962, in \textit {Atomic and Molecular Processes}, ed. D. R. Bates (Academic Press, New York)
\bibitem[\protect\citeauthoryear{Author}{2013}]{}Walmsley, C. M., Churchwell, E., Nash, A., Fitzpatrick, E. 1982, ApJ, 258, L75
\end{thebibliography}

\label{lastpage}
\end{document}